\documentclass[%
reprint,
 amsmath,amssymb,
 aps,
]{revtex4-1}

\usepackage{graphicx}
\usepackage{dcolumn}
\usepackage{bm}

\begin{document}

\title{Measurement and Data-Assisted Simulation of Bit Error Rate in RQL Circuits}

\author{Quentin Herr, Alex Braun, Andrew Brownfield, Ed Rudman, Dan
  Dosch, Trent Josephsen, and Anna Herr}
\affiliation{Northrop Grumman Corp., Baltimore, MD 21240}

\date{14 April 2021}

\begin{abstract}
A circuit-simulation-based method is used to determine the
thermally-induced bit error rate of superconducting logic
circuits. Simulations are used to evaluate the multidimensional
Gaussian integral across noise current sources attached to the active
devices. The method is data-assisted and has predictive
power. Measurement determines the value of a single parameter,
effective noise bandwidth, for each error mechanism. The errors in the
distributed networks of comparator-free RQL logic nucleate across
multiple Josephson junctions, so the effective critical current is
about three times that of the individual devices. The effective noise
bandwidth is only 6-23\% of the junction plasma frequency at a modest
clock rate of 3.4\,GHz, which is 1\% of the plasma frequency. This
analysis shows the ways measured bit error rate comes out so much
lower than simplistic estimates based on isolated devices.
\end{abstract}

\maketitle

Digital superconducting is a long-standing candidate
for beyond-CMOS technology due to high clock rates and unparalleled
power efficiency
\cite{herr2011ultra}-\nocite{holmes2013energy}\cite{vernik2016energy}.
Demonstrations of digital functions continue to mature
\cite{herr20138}, and are currently at the level of small CPUs
\cite{ayala2021mana}. Fundamental power advantages derive from both
the Josephson junction active devices and non-dissipative
interconnects. In contrast to CMOS, superconducting technology is
thermally limited not device limited, meaning the devices are sized
based on bit-error rate (BER) limitations, not based on lithographic
minimum feature size. Current energy-efficient variants of
superconducting digital logic, such as RQL and the QFP, have
scaled the energy per switching event to within a factor of 100-1000
above Landauer's limit, $\ln(2)k_BT$.

The penalty for such performance is more stringent optimization
criteria. The circuits need to be characterized not only by timing
design, as with CMOS, but also in terms of parametric operating
margins and BER performance. Parametric margins are determined using
noise-free circuit simulations, but the more relevant question is to
determine parametric margins defined in terms of an acceptable
BER. For large scale applications, the acceptable BER for individual
gates is on order $10^{-24}$. These levels are difficult to access in
either simulation or measurement.

Elevated BER can be measured as a function of a tuning parameter, and
extrapolated to the region of unobservable levels. This approach has
been widely reported \cite{herr1996error}-\nocite{rylyakov1999pulse,
  bunyk2001experimental, fujiwara2005error, wetzstein2011comparison,
  tanaka2013bit}\cite{takeuchi2017measurement}. The usual caveats with
extrapolation apply, as there are multiple sources of BER in the
circuits. Regardless of the error mechanism, BER scales simply with
energy set by the device size \cite{herr2011ultra}. While measurement
is useful and necessary, a simulation-based approach is also needed.

Models of BER exist for only a few simple cases. Analysis of the
comparator \cite{filippov1995signal} predicts how the BER scales with
clock rate in the low-speed limit that might be applicable to the QFP
\cite{takeuchi2017measurement}, but not more generally. Spontaneous
switching of biased, isolated junctions is understood
\cite{klein1982thermal}, but has not been generalized to the
distributed networks of SFQ circuits.

Simulation methods are also of limited utility. Fokker-Plank
simulation has been applied to the comparator \cite{herr1997error}, but
this approach is prohibitive for larger systems. The only alternative
is a Monte-Carlo method requiring repeated circuit simulations with
random sources for the noise \cite{satchell1999limitations}. The
method is generally applicable to any circuit schematic, but is low
resolution and numerically intensive.

We present a new simulation-based method that is a data-assisted
multivariate integration.  The method repurposes the integral for
parametric yield to find BER:
\begin{itemize}
  \item
  Parametric Yield is the integral of the Gaussian-distributed
  parameters over the operating region of the circuit. {\it This gives
    the probability of working circuits.}
  \item
  Bit Error Rate is the integral of the Gaussian-distributed noise
  currents over the operating region of the circuit. {\it This gives
    the probability of working clock cycles.}
\end{itemize}
A single free parameter, effective noise bandwidth, is determined for
each scenario by comparing simulation to measurement. The approach is
generally applicable to any circuit schematic and any noise-induced
error mechanism associated with Single-Flux-Quantum (SFQ)
circuits. The method is deterministic and can calculate arbitrarily
small error probabilities with high resolution. The method is
efficient in terms of circuit simulations, requiring about $3^N$
margin calculations for $N$ noise sources. Noise sources are included
no more than eight at a time, but a sliding window of inclusion covers
larger circuits.

BER simulation can be used to improve parametric optimization in the
design phase. An important feature here is to ability to calculate
individual margins (and BER) as an intersection across global
parametric corners. At each corner parameter set, margins are
calculated in parallel, and the final number is the most pessimistic
result. This approach could be used to calculate the most pessimistic
BER across global parameter mistargeting, or across signal input
timing. This is adjacent to the statistical timing analysis of
advanced-node CMOS, see e.g. \cite{kahng2015new}.

The rest of the paper covers simulation-based numerical evaluation of
the multivariate Gaussian integral, simulated BER of two
representative logic gates, and measured BER and comparison to the
simulation. The data-assisted method uses a generic value for
effective bandwidth in the initial simulations, and then solves for
the measured values. The discussion centers on the physical meaning of
the effective noise bandwidth.

\section{Multivariate Gaussian Integral}

BER is the probability of failure, which is given by the complement of
yield.  This is the integral over the tails of the Gaussian distribution,
outside the operating region of the circuit.
We start with the one dimensional case and generalize to higher dimensions.
In the simplest case the upper and lower margins of the circuit are equal. 
Normalized margin $r=m/\sigma$ is the circuit operating margin
$m$, normalized in units of the standard deviation. 
In this case
\[
  \mbox{yieldc}(r)=\mbox{erfc}\!\left(r/\sqrt{2}\right)
\]
The upper and lower margins, $r_a$ and $r_b$, are likely different. If
so, the contributions can be computed separately, normalized, and
summed.
\[
  \mbox{yieldc}(1,r_a,r_b)=1/2\,\mbox{erfc}\!\left(r_a/\sqrt{2}\right)
  +1/2\,\mbox{erfc}\!\left(r_b/\sqrt{2}\right)
\]
In $N$ dimensions, analytic results obtain if the operating region is
defined by a radius $r$.
\[
  \mbox{yieldc}(N,r)=\mbox{Q}(N/2,r^2/2)
\]
where $\mbox{Q}(a,x)=\Gamma(a,x)/\Gamma(a)$, is the normalized upper
incomplete gamma function as defined e.g. in
\cite{pressnumerical62}. This is also known as the regularized gamma
function. Here the limits of integration correspond to a constant
value of the Gaussian joint probability distribution, which has radial
symmetry.

\begin{figure}
 \centering
 \includegraphics[width=1.5in]{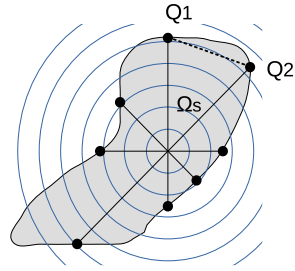}
 \caption{The operating region of the circuit, shaded, overlays the
   Gaussian joint probability distribution, shown as concentric
   isocontours equally spaced in units of sigma. The operating region
   is approximated by points on the operating boundary, organized as
   simplexes (line segments in the two dimensions shown). The simplex
   defined by points Q1 and Q2 subtends an angle $\Omega_s$. The mean
   value of the Q-function for these points, weighted by the angle, is
   one term in the Riemann sum that approximates the Gaussian integral
   over the operating region.
  \label{riemann}}
\end{figure}

The operating region of the circuit generally does not have radial
symmetry and must be mapped out. This is illustrated for two
dimensions in Fig.~\ref{riemann}, and generalizes to higher
dimensions. We approximate the operating region by connecting boundary
points with simplexes. Each simplex $s$ gives
a yield estimate based on the mean value $\langle\mbox{Q}_s\rangle$ of
the $Q$ values evaluated at the boundary points. This estimate is
weighted by the angle $\Omega_s$ subtended by the segment. The
integral is approximated by a normalized Riemann sum of these
terms. In $N$ dimensions,
\[
  \mbox{yieldc}=\frac{1}{\Omega_N}\sum_s\Omega_s\langle\mbox{Q}_s\rangle
\]
which is normalized to the total angle $\Omega_N$. The mean value of
Q for each simplex is
\[
  \langle\mbox{Q}_s\rangle=\frac{1}{N}\sum_{i=1}^N\mbox{Q}(N/2,r_{t[s,i]}^2/2)
\]
Simplexes are indexed by $s$. For a given simplex, $r_{t[s,i]}$
returns the value of radius, indexed by $i$, for each of the points.

Details of the algorithm are described in Appendix 1. Including
calculation of the angle $\Omega_s$, the annealing schedule of
the adaptive algorithm that adds new points to the operating boundary,
and simplex fracturing to create a finer grid.

The algorithm does not require the operating region to be convex,
which is a fundamental constraint in simplicial design
centering \cite{director1977simplicial, herr2001improved}. Here it is
only required that the operating region be single-valued as viewed
from the origin, which corresponds to the binary search vectors. The
value of the integral is dominated by regions of the operating
boundary nearest the origin. This implies that the computational
intensity is similar to finding global minima by exhaustive search.
In practice, covering the space exhaustively requires about $3^N$
margin calculations. This can be understood as the set of all search
vectors for which each dimension has a value of $+1$, $-1$, or
$0$. This set captures all of the potential dependencies of the space
and of all subspaces. The number of simplexes grows rapidly as points
are added, which currently imposes a practical limit of about eight
dimensions.

\section{BER Simulation}

In simulation, an auxiliary dc current source is applied across each
junction in the circuit. The current through each source is a circuit
parameter, with nominal value set to zero and sigma equal to rms noise
current,
\[
I_{\mbox{{\small rms}}}=\sqrt\frac{4k_BTI_c\beta}{\Phi_0}
\]
which scales with junction critical current, $I_c$. The expression
derives from the Johnson noise, $\sqrt{4k_BTB/R}$ using the junction
characteristic impedance, $R=\sqrt{L_J/C_J}$, plasma frequency,
$B=1/(2\pi\sqrt{L_JC_J})$, and inductance, $L_J=\Phi_0/(2\pi I_c)$.
The bandwidth $\beta$ is a unitless fraction of the plasma frequency.
We will use the effective bandwidth $\beta$ as a fitting parameter to
relate the simulation result to measured data.  This expression for
noise current does not depend on junction capacitance, $C_J$ or shunt
resistance.

 The overall method is 1) choose an initial value for $\beta$ to
 simulate BER, 2) compare the simulated result to measured data to
 find the experimentally-derived effective $\beta$ for each error
 regime, and 3) use this value in subsequent simulations. All BER
 simulations in this paper used the initial value $\beta=0.25$.

On the way to evaluating the Gaussian integral, the algorithm reports
the critical search vector, directed towards the point on the
operating boundary found to be nearest the origin. This indicates
which sources contributed most to nucleating the error. The critical
vector is suggestive but not definitive. A gate e.g. with symmetric
inputs would have equally critical vectors on both, but only one could
be reported. However, the algorithm would integrate over all paths to
calcualte BER.

It is intractable and unnecessary to include all noise sources at once
in the BER simulation. Instead, it is enough to include sources that
contribute to nucleation of the error.  Simulations proceed with
inclusion 7-8 noise-current parameters at a time, chosen among
contiguous junctions. We use a sliding window, meaning that the
parameter-inclusion sets overlap. Each window returns a BER. The most
relevant window is the one that returns the highest BER.

\begin{table*}
\renewcommand{\arraystretch}{1.0}
\caption{Critical vectors for the points indicated in the AND gate
  simulation of Fig.~\ref{andsim}a, illustrating three different error
  mechanisms. Columns correspond to noise sources on the junctions of
  Fig.~\ref{gates}a}
\label{window}
\centering
\begin{tabular}{cc|rrr|rrr|rrrrr}
  \multicolumn{1}{c}{} &
  \multicolumn{1}{c|}{} &
  \multicolumn{3}{c|}{Input A} &
  \multicolumn{3}{c|}{Input B} &
  \multicolumn{5}{c}{Output} \\
  \multicolumn{1}{c}{Point} &
  \multicolumn{1}{c|}{Window} &
  \multicolumn{1}{c}{J0} &
  \multicolumn{1}{c}{J1} &
  \multicolumn{1}{c|}{J2} &
  \multicolumn{1}{c}{J3} &
  \multicolumn{1}{c}{J4} &
  \multicolumn{1}{c|}{J5} &
  \multicolumn{1}{c}{J6} &
  \multicolumn{1}{c}{J7} &
  \multicolumn{1}{c}{J8} &
  \multicolumn{1}{c}{J9} &
  \multicolumn{1}{c}{J10} \\
  \hline
  a & 1  & $-$0.271 & $-$0.924 & $-$0.271 &    0.000 & 0.000 &    0.000 &    0.000 &    0.000 & -     & -     & -     \\
  b & 1  &    0.197 &    0.558 &   0.197  &    0.279 & 0.674 &    0.279 &    0.000 &    0.000 & -     & -     & -     \\
  c & 2  & -        & -        &   0.163  & -        & -     &    0.000 &    0.543 &    0.774 & 0.231 & 0.163 & 0.000 \\
\hline
\end{tabular}
\end{table*}

\begin{figure}
 \centering
 \includegraphics[width=3.4in]{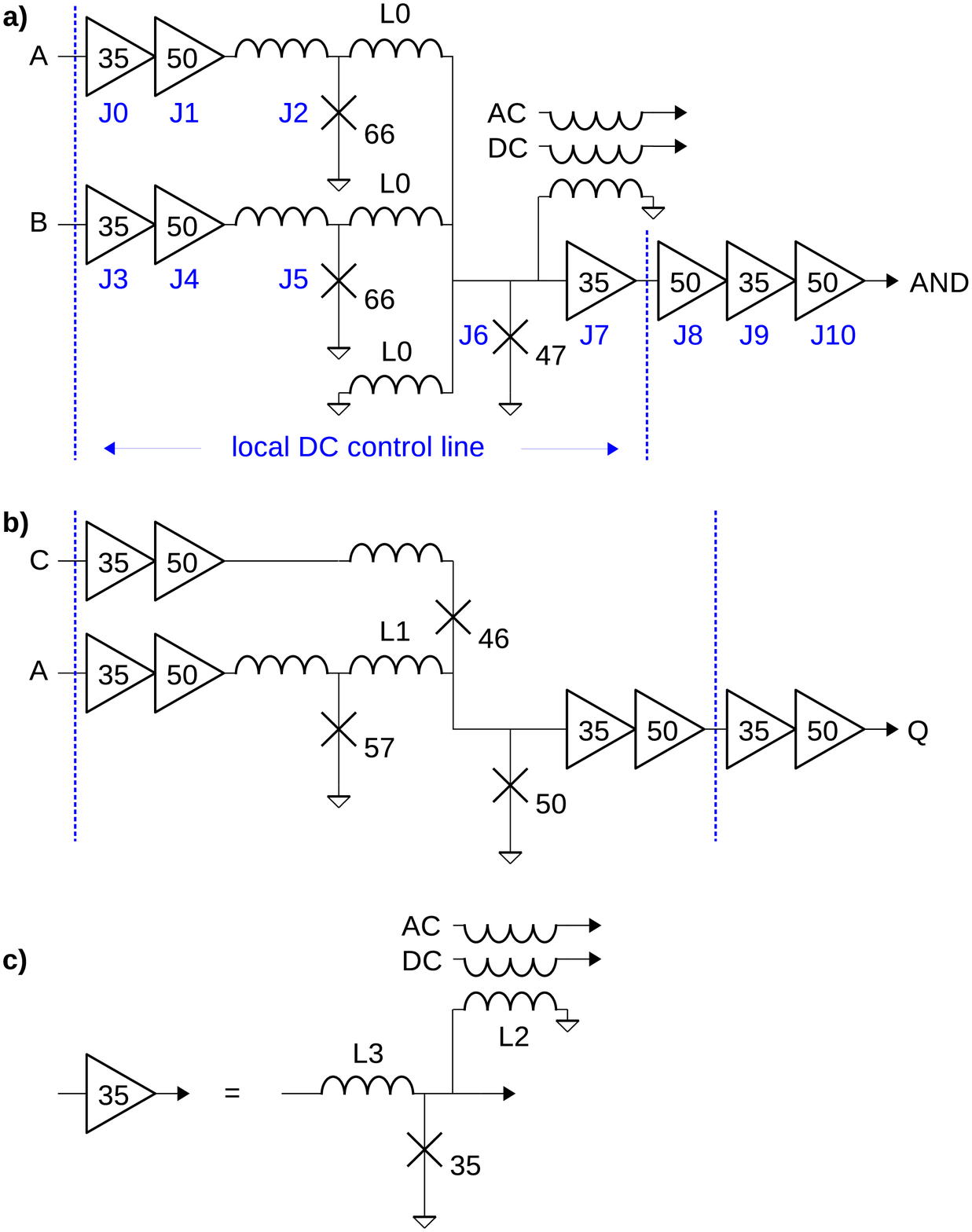}
 \caption{Circuit schematics are shown for a) the AND gate, b) the DFF
   gate, and c) the JTL subcircuit. The critical currents of the
   junctions, $I_c$ are given in units of $\mu$A. JTL junctions are
   biased with an AC sinusoid and a DC offset via mutual inductance to
   L2. The signal inductors, including L3, typically have values
   ranging from 0.33-0.5\,$\Phi_0/I_c$. The storage inductors L0 and
   L1, and the bias inductors L2, are larger. The AC bias is applied
   globally, including the input and output amplifier stages. The DC
   bias is applied locally to the device under test via a separate
   control line.
  \label{gates}}
\end{figure}

We applied the algorithm to the RQL circuits shown in
Fig.~\ref{gates}. For present purposes, the AND gate is representative
of gate implementations that do not have a Josephson comparator, and
the DFF serves as a diagnostic for comparator-based gates, including
RSFQ and the QFP. The AND gate \cite{braun2018rql} is based on
three-input majority, with one input tied to ground. This is analogous
to QFP logic, but implementation in terms of data encoding,
interconnect, and clock phasing are distinct. The RQL implementation
is low-latency as it does not require advance of the clock phase to
prevent back-traveling pulses on the inputs
\cite{braun2018superconducting}; multiple gates can be cascaded on a
single phase. The D flip-flip (DFF) stores the input until it is
clocked out through the comparator. The implementation is similar to
the RSFQ DRO. However, both positive and negative polarity pulses are
used in RQL, so the stored signal may be of either polarity. The
utility of the DFF and other features of the RQL library are beyond
the scope of BER considerations, and will be described elsewhere.

\begin{figure}
 \centering
 \includegraphics[width=3.72in]{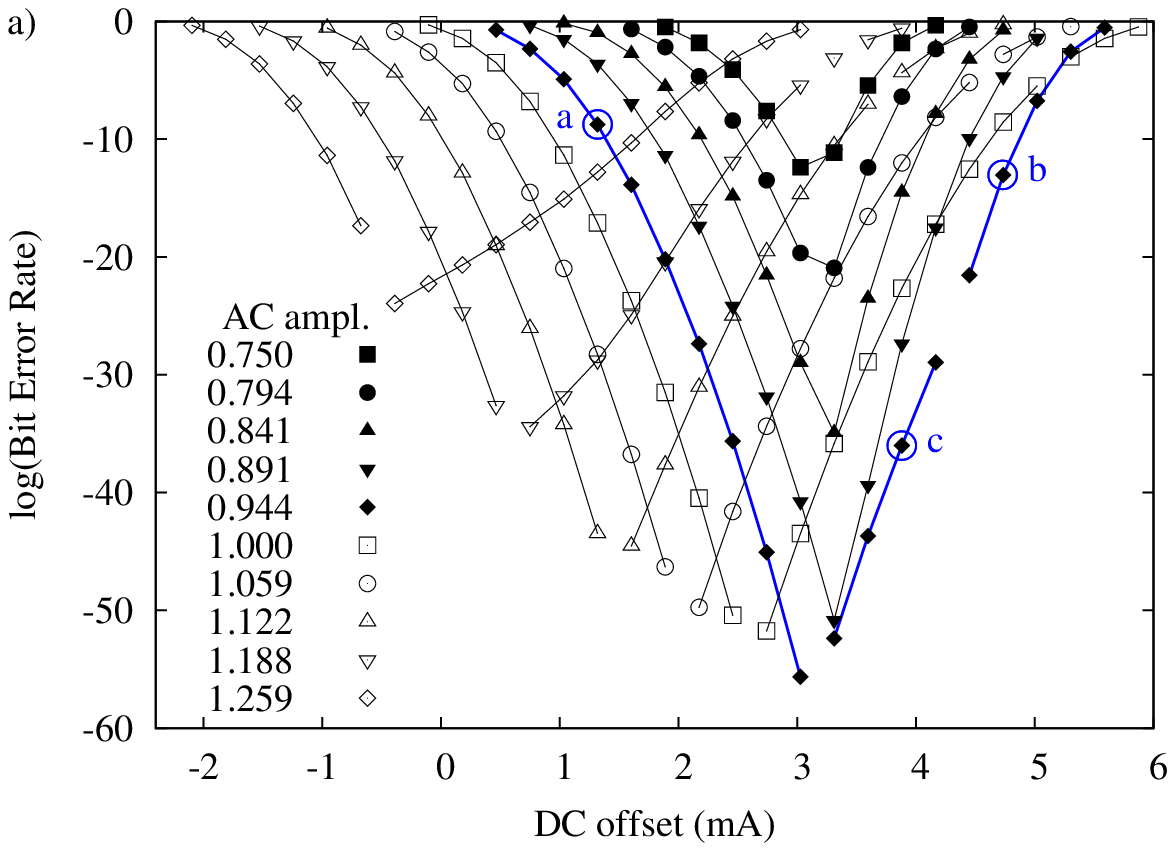}
 \includegraphics[width=3.72in]{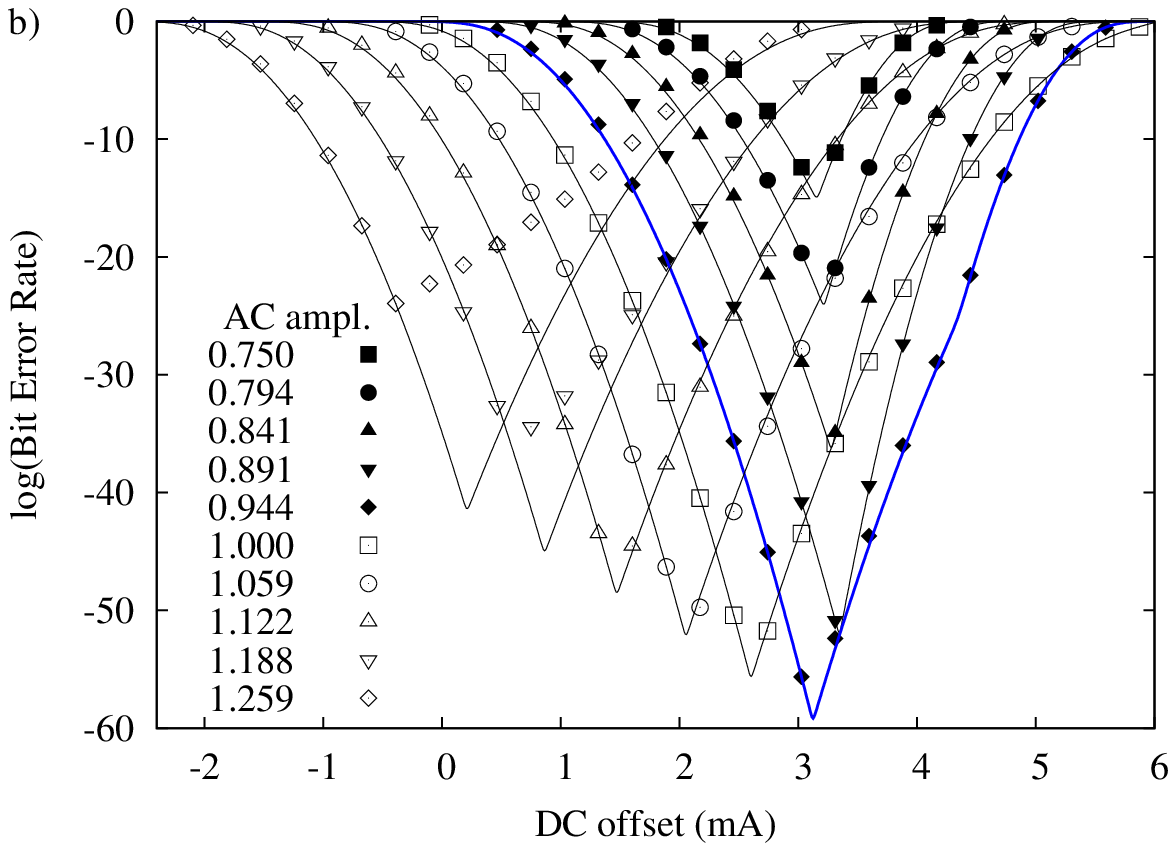}
 \includegraphics[width=3.72in]{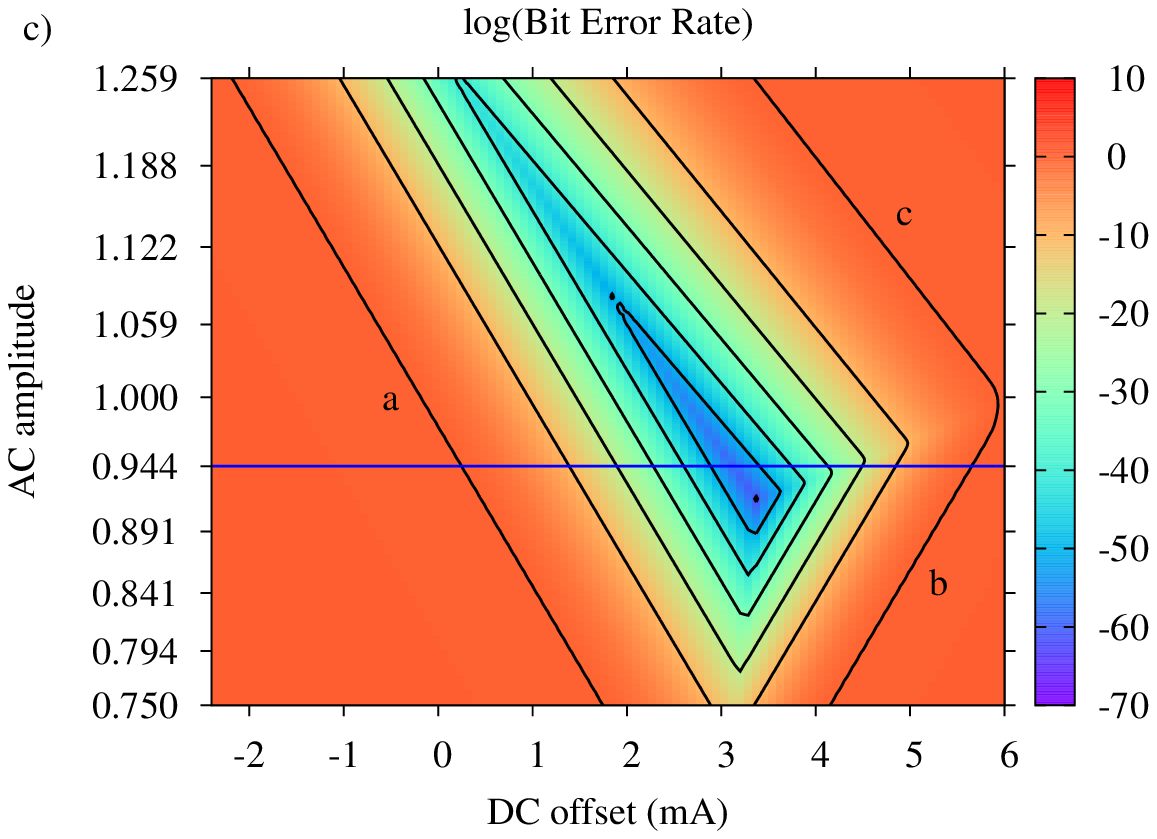}
 \caption{Simulated BER of the AND gate as a function of DC level,
   $x$, and normalized AC amplitude, $A$. a) Simulations (points) for
   each value of $A$ are connected with lines to guide the eye. Gaps
   in the line are where the maximum BER moved from one sliding window
   to the other. Two sliding windows were used for inclusion of the
   noise current sources. The points (a), (b), and (c) are called out
   for further analysis. b) Simulated points replotted, now with
   curves defined by Eqn.~\ref{erfcsim} that captures the result in
   functional form. c) The function of Eqn.~\ref{erfcsim} plotted as a
   contour map. The outer contour is for a BER of 0.5, which equates
   to noise-free margins. The horizontal blue line marks a
   cross-section corresponding to the blue line in the upper
   panels. Each contour has three sides (a), (b), and (c) representing
   three distinct error mechanisms.
  \label{andsim}}
\end{figure}

The BER calculation at each set point requires a few thousand binary
searches across multiple windows. However, simulations can proceed in
parallel, so full characterization of the gate requires less than
24\,hours of real time using a few large servers running WRspice
\cite{wrspice}.

Simulated BER of the AND gate is shown Fig.~\ref{andsim}a. Three sets
of curves are visible; the progression with negative slope in the
left, and two progressions of positive slope on the right. All of this
can be expressed analytically as the sum of three error functions,
with fitting coefficients that depend on the AC amplitude, $A$. The
fit, detailed below, produced the contour plot shown in
Fig.~\ref{andsim}c.

The three-sided contours correspond to typical JTL-based error
mechanisms. The DC offset optimally produces symmetry between positive
and negative SFQ pulses by putting a $\Phi_0/2$ flux bias on the bias
inductor (L2 in Fig.~\ref{gates}c). AC amplitude powers the
transistions. When DC is low (side a), positive pulses fail to
propagate and are annhiliated by the trailing negative pulse. When DC
is high and AC is low (side b), it is the trailing negative pulse that
fails to propagate. When DC and AC are both high (side c), positive
pulses are spontaneously generated in error. By symmetry, one might
expect that DC low and AC high would produce a fourth regime where
negative pulses were generated in error. The asymmetry is in the data
encoding. Every positive pulse is followed half a cycle later by a
negative pulse that serves as a reset. Spontaneous negative pulses
that anticipate the transistion do not register as an errors. This
extends the operating region of the circuit up and to the left well
beyond normal design limits.

In the AND gate, the three error mechanisms may arise in different
parts of the circuit. This is illustrated for the three points (a),
(b), and (c) called out in Fig.~\ref{andsim}a. The critical vectors
for these points are entered in Table~\ref{window}. Point (a) is
centered on one of the input JTLs, point (b) straddles both inputs,
and point (c) is centered on the output.  For each case, the table
shows the dominant sliding window of parameter inclusion.

It would be possible---but needlessly intensive---to run and rerun the
BER simulations for different values of effective bandwidth in order
to match the experimentally measured result. Instead, we develop a
functional fit to the simulations that incorporates the bandwidth
dependance. For each value of $A$, the simulation points were
partitioned and fit to one of three error functions, using margin and
noise coefficients. The total BER is given as the sum of these three
terms.
\begin{align}
  \mbox{BER}(x,A,\beta_a,\beta_b,\beta_c)=
  &\frac{1}{2}\mbox{erfc}\!\left[\frac{x-m_a(A)}{\sqrt{2\beta_a}n_a(A)}\right]
  \nonumber \\
  +  \frac{1}{2}\mbox{erfc}\!\left[\frac{m_b(A)-x}{\sqrt{2\beta_b}n_b(A)}\right]
  + &\frac{1}{2}\mbox{erfc}\!\left[\frac{m_c(A)-x}{\sqrt{2\beta_c}n_c(A)}\right]
    \label{erfcber}
\end{align}
The three similar terms fit the (a), (b) and (c) curves. 
For the first term, the fitting parameter $m_a(A)$,
corresponds to noise-free margin which determines the offset along the
$x$-axis. The term $\sqrt{\beta_a}n_a(A)$ corresponds to rms noise,
with the dependance on effective noise bandwidth, $\beta_a$, made
explicit. All of
the curves are well-characterized by a single function using fitting
coefficients that are linear functions of $A$.
\[
  m_a(A)=Am_{1a}+m_{0a} \mbox{,}\quad n_a(A)=An_{1a}+n_{0a}
\]

\begin{table}
\renewcommand{\arraystretch}{1.3}
\caption{Coefficients for the function of Eqn.~\ref{erfcber} fit to
  the AND gate simulation points.}
\label{coef}
\centering
\begin{tabular}{l|rr|rr}
    & $m_{1\_} $ & $m_{0\_} $ & $n_{1\_} $ & $n_{0\_}$ \\
\hline
  a &  $-$7.724 &    7.542 &  0      & 0.1760 \\
  b &     7.732 & $-$1.650 &  0      & 0.1240 \\
  c & $-$10.173 &   16.151 &  0.0655 & 0.1480 \\
\hline
\end{tabular}
\end{table}

Fit values for the three curves are entered in Table~\ref{coef}. Note
that the slopes of curves (a) and (b) are constant across the range of
$A$ values. The resulting function
\begin{equation}
  \mbox{BER}(
  x,
  A,
  \beta_a\mbox{=}0.25,
  \beta_b\mbox{=}0.25,
  \beta_c\mbox{=}0.25)
    \label{erfcsim}
\end{equation}
is a good fit to all of the simulation points, as shown in
Fig.~\ref{andsim}b. The exception is that the positive-slope
progressions for $A=\{1.188, 1.259\}$ are above the line. Further
investigation showed that these were over-bias errors in the output
JTL, far from the gate itself. These errors are real, but are outside of
the gate characterization.

\begin{figure}
 \centering
 \includegraphics[width=3.72in]{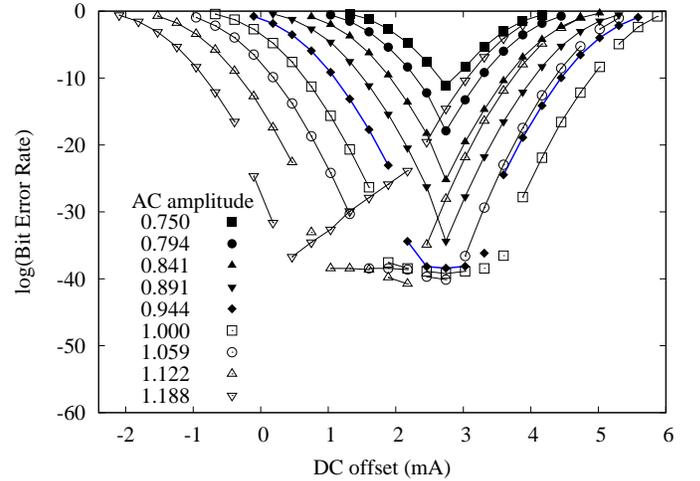}
 \caption{Simulated BER for the DFF gate as a function of the local DC
   level and normalized AC amplitude, $A$. Four sliding windows were
   used, centered on each of the inputs, the output, and the gate
   itself. Points for each value of $A$ are connected with lines. Gaps
   in the line are where the maximum BER moved from one window to the
   other. All four windows are represented in the data for $A=0.944$
   (bold lines). The noise floor arises from switching errors in the
   gate comparator.
  \label{dff}}
\end{figure}

We applied the same method to BER simulation of the DFF, shown in
Fig.~\ref{dff}. Here the waterfall curves are interrupted by a noise
floor at about $10^{-40}$. Further analysis showed that the floor
corresponds to decision errors in the comparator, while the waterfalls
corresponds to errors in the JTLs surrounding the gate. The floor is
far below the limits of direct observation in measurement. This
simulation indicates that the AC amplitude and DC offset set points,
applied directly to the JTLs but not to the comparator itself, do not
allow exapolation to the dominant error mechanism at low BER. What is
needed is an auxillary current bias applied directly to the central
node of the comparator \cite{herr1996error, filippov1995signal}, as
this would produce a large shift in comparator threshold.

\section{BER Measurement and Comparison to Simulation}
Integrated circuits containing the RQL gate library were designed into
a six-metal-layer fabrication process supplied by D-Wave
\cite{berkley2010scalable}. This revision of the fab featured
Josephson junctions with 100\,$\mu$A/$\mu$m$^2$ critical current
density. The logic gates used individual circuits with dedicated
input stages and output amplifiers on-chip. The circuits shared a
global 3.4\,GHz resonant clock network that provisioned the entire
active area of the chip.

Crosstalk between the clock and signal lines in the pressure-contact
LHe dip probe was a challenge, so the output data link was established
using differential signals, amplified by a pair of Miteq
JSMF4-02K180-30-10P LNAs at the probe head. 3\,dB attenuators
(50\,$\Omega$) at the inputs of the LNAs were used to improve the
impedance match. These fed a Marki BAL-0026 balun that produced a
single-ended signal for the lab instruments. A Keysight J-BERT M8020A
produced the data patterns and counted the errors. This was
synchronized to a pair of Rohde\,\&\,Schwarz SGS100A sources that
powered the clock network with I and Q signals. DC levels were
generated by a Stahl BS 1-10.

\begin{figure}
 \centering
 \includegraphics[width=3.72in]{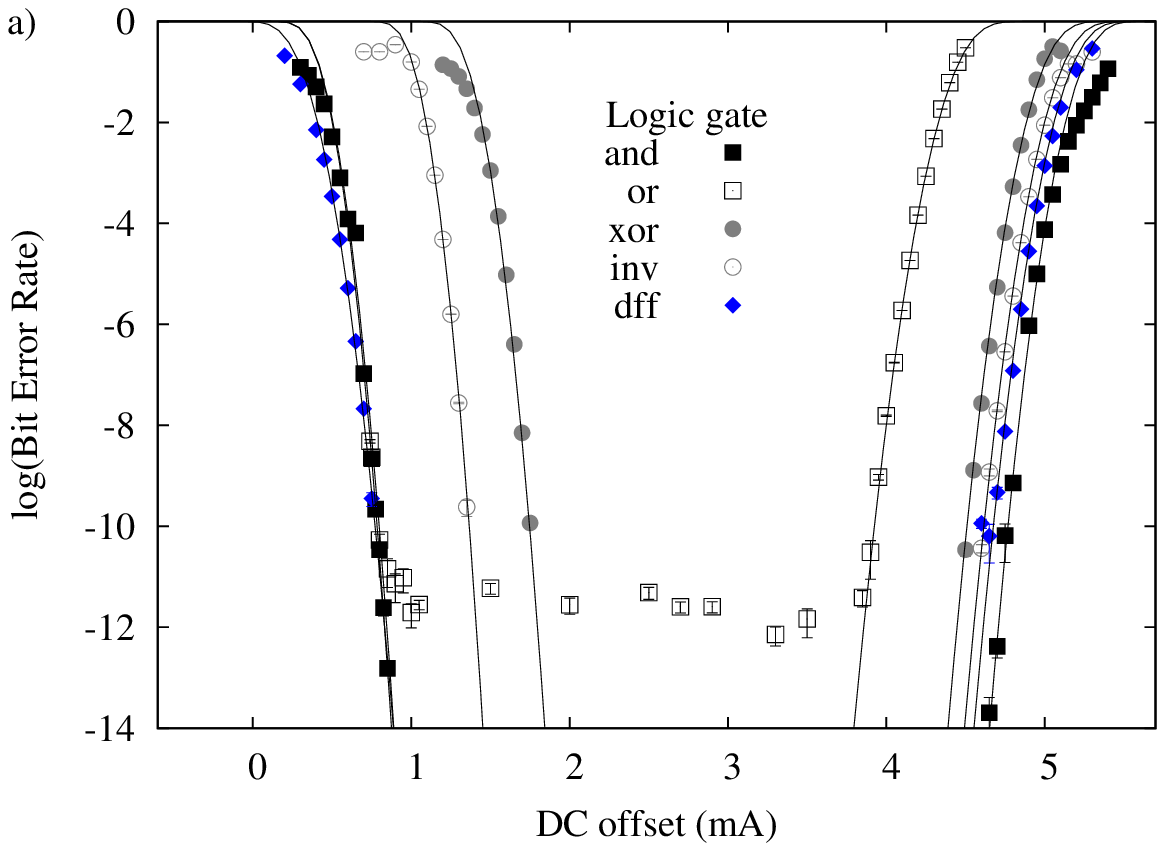}
 \includegraphics[width=3.72in]{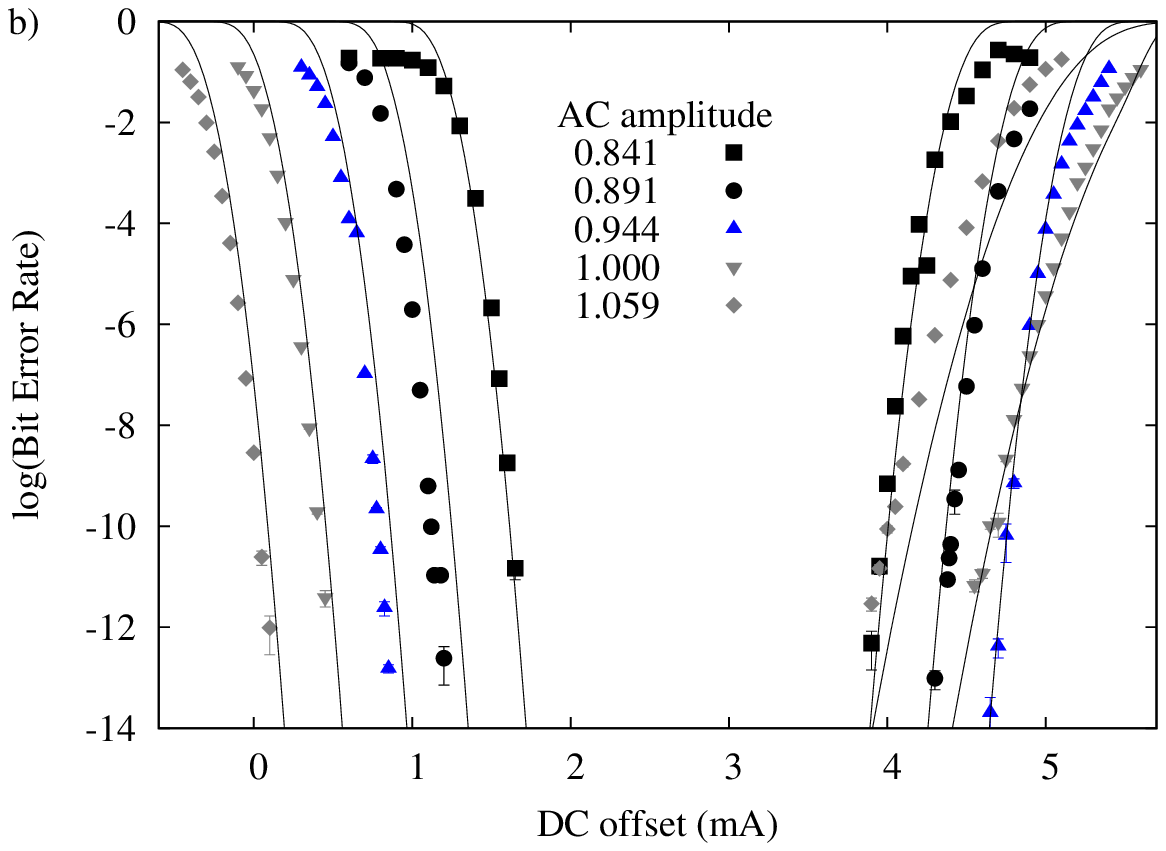}
 \includegraphics[width=3.72in]{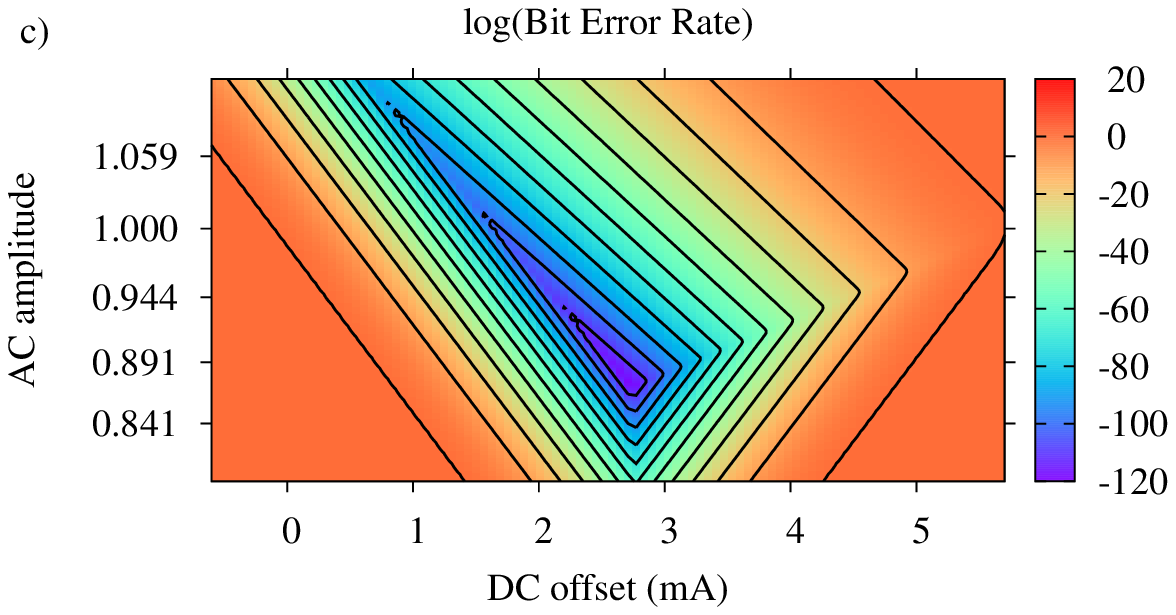}
 \caption{Measured BER (points) as a function of DC offset. Error bars
   visible for the lowest points correspond to binomial counting
   statistics. a) Five different logic gates measured at an AC
   amplitude, $A=0.944$. Each curve is fit to an error function to
   guide the eye. A noise floor is visible for the OR gate. We
   attribute this to observed limitations of the output data link for
   this particular case, and not to the gate itself. b) Additonal
   measurements of the AND gate at several values of $A$. Measurements
   are fit to the simulation result (lines) via Eqn.~\ref{erfcfit}. c)
   The function of Eqn.~\ref{erfcfit} plotted as a contour map. The
   outside contour is for a BER of 0.5, corresponding to noise-free
   margins. The inside contour is for $0.5\times 10^{-110}$.
  \label{data}}
\end{figure}

Measurement of five different logic gates is shown in
Fig.~\ref{data}a. The AND and OR gates are closely related as they
derive from the same majority gate. For the OR, the third input is
initialized to logical ``1.'' The Inverter and XOR
\cite{harms2019superconducting} are also closely related; the Inverter
is derived from XOR, with one of the inputs initialized to logical
``1.'' BER for AND and OR are coincident on the left, but offset on
the right. BER for the Inverter and XOR are nearly coincident on the
right, but offset on the left. These offsets might be inherent to the
design and test vectors, or due to parametric variations in fab or
trapped-flux effects in test. Additional simulations and test would
be needed to resolve these questions. The main result is that the
slopes of the curves are consistent, reflecting that the observed
errors are induced in interconnect JTLs, which are similar among the
designs.

Measured BER of the AND gate at different AC and DC set points, shown
in Fig.~\ref{data}b, mirrored the simulations. Curves fitting the
simulation to the measurement, also shown, used the function defined
in Eqn.~\ref{erfcber} with arguments
\begin{equation}
  \mbox{BER}(
  x\mbox{=}1.04x',
  A\mbox{=}0.99A',
  \beta_a\mbox{=}0.058,
  \beta_b\mbox{=}0.14,
  \beta_c\mbox{=}0.23)
    \label{erfcfit}
\end{equation}
where $x'$ and $A'$ are the DC offset and AC amplitude scale for the
measured data. These numbers produce good quantitative agreement. The
factor of 1.04 indicates a small registration error on DC offset. This
could be accounted for by a 4\% shift in mutual inductance to the
control line. The AC amplitudes for the measurement have an arbitrary
scale, relative to simulation, but the spacing of 0.5\,dB is fixed for
both. A single scaling factor, 0.99, aligned the simulation to the
measurement. These are small disparities given that the simulation
used nominal design values; no attempt was made to incorporate global
parametric targeting. The primary result lies in the values for
effective noise bandwidth. These numbers were chosen to be 0.25 in
simulation, but came out widely divergent---and generally smaller---in
measurement.

Measured BER of the DFF were similar to the AND gate. This is
unsurprising, as simulation indicates that observable errors belong to
the JTLs, not the gate itself. However, the DFF comparator produced a
BER floor in simulation. To induce comparator errors at observable
levels, an additional detuning bias would be needed that directly
shifts the comparator threshold. We would expect the effective noise
bandwidth for comparator errors to be higher than the other error
mechanisms, as the error does not depend on junction switching time
per se, but on a shorter time scale when the phase difference between
the junctions passes some threshold. Based on a review of previous
work \cite{herr1997error, filippov1995signal}, an effective normalized
bandwidth of 0.5 is our best estimate. If so, the simulations using a
bandwidth of 0.25 underestimated comparator errors, and the BER floor
is actually around $10^{-25}$ based on scaling given in
Eqn.~\ref{erfcber}.

\section{Discussion}

Results of the previous section show that 1) operating margins at low
BER are generally wide enough to support complex RQL circuits. 2) BER
in measurement is less than in simulation, if effective noise
bandwidth is taken to be the Josephson junction plasma frequency, and
3) the effective noise bandwidth depends on the error mechanism.

AC overbias errors, corresponding to curves (c) in Figs.~\ref{andsim}
and \ref{data}, involve spontaneous junction triggering that can be
modeled using the Kramers escape rate. This analysis gives a
first-principles estimate without resort to an effective noise
bandwidth. The Kramers escape rate is $f\approx f_a\exp\left[-\Delta
  U/(k_BT)\right]$, where $k_BT$ is the thermal energy, $\Delta U$ is
the potential barrier, and $f_a$ is the attempt frequency.  Here we
can safely model the noise as purely thermal at 4.2\,K, because DC and
AC currents are very weakly coupled to the circuit using magnetic
transformers.  External noise scales as $k^2$, where $k$ is coupling
coefficient \cite{herr2011ultra}.

The potential barrier for a single biased Josephson junction is
approximately \cite{likharev1986dynamics}
\[
\Delta U(I_b) \simeq\frac{\Phi_0}{2\pi} \frac{2I_c}{3}
\left(2 -\frac{2I_b}{I_c} \right)^{3/2}
\]
where the total bias current through the junction is the sum of AC and
DC components, $I_b=I_{\mbox{DC}}+I_{\mbox{AC}}$. At the nominal value
$I_{\mbox{DC}}=2.8$\,mA, the externally-applied DC offset current
produces a bias current of only $I_b=0.24 I_c$ through the
junction. This gives the transfer function $I_b/I_c=0.086
x/\mbox{mA}$. The AC current $I_{\mbox{AC}}$ shifts the BER function
by an arbitrary offset---for comparing to measured data this is left
as a fitting parameter. The resulting Kramers escape rate is
\begin{equation}
  \mbox{BER}(x) \simeq\alpha\exp\!\left[
  \frac{\Phi_0 I_c}{3\pi k_BT} (2\! - \! 2[0.086 I_{DC}\! + \! I_{AC}]/\mbox{mA})^{3/2} \right]
    \label{kramers}
\end{equation}
where the prefactor $\alpha=10$ is the number of attempts per clock
cycle. This value does not need to be precise for low BER values. The
function corresponds to the asymptotic form of erfc, and is valid in
the region of low BER. Fitting the measured data for overbias errors
(the steeper curves on the right of Fig.~\ref{data}b) with
Eqn.~\ref{kramers}, using critical current and AC bias as fitting
parameters, results in $I_c\approx 110\,\mu$A.

The effective critical current indicated by the Kramers escape rate is
about three times larger than the critical current of the individual
junctions used in a circuit, suggesting that the error mechanism is
distributed across multiple junctions.  This result comports nicely
with the simulation result shown as curve (c) in Table~\ref{window}
with about three junctions involved in the nucleation of overbias
errors in the output JTL. The JTL inductances have an $I_cL$ product
of about $\Phi_0/3$, meaning that the three JTL junctions span about
one-$\Phi_0$ of interconnect.

The normalized effective bandwidth for overbias errors in measurement,
$\beta_c=0.23$, agrees well with initial simulations with
$\beta_c=0.25$ and indicates a spontaneous-switching time that is long
compared to the plasma frequency. This simply means that junction
switching time is longer at low values of overbias, as observed in
\cite{harris1979turn}.

Annihilation errors, corresponding to curves (a) in
Figs.~\ref{andsim} and \ref{data}, have a different character. These
errors arise when the leading, positive pulse in the RQL data encoding
fails to propagate and is annihilated by the trailing negative
pulse. Since the positive and negative pulses are separated in space
and time by a half-cycle, the noise must persist for a fair fraction
of a clock cycle in order to produce an error. This implies much lower
effective noise bandwidth. The measured value, $\beta_a=0.058$
corresponds to about 17\% of the clock cycle, or 67\% of the
quarter-cycle of the four phase clock at 3.4\,GHz clock and 340\,GHz
plasma frequency. Since noise bandwidth for annihilation errors is
determined by clock rate it is expected to increase with clock rate,
to be confirmed in future experiments.

Similar arguments exist for annihilation error corresponding to curve
(b) in Fig. 3-5, but here the trailing, negative pulse of the RQL data
encoding fails to propagate and is annihilated when there is a
positive pulse on the next cycle. Different effective bandwidths
between curves (a) and (b) is unsurprising as the data encoding is
asymmetric, and as the errors arise in different parts of the circuit,
based on the critical vectors entered in Table~\ref{window}.

\section*{Conclusion}

Here we have presented a framework to explain, predict, and quantify
the BER of RQL gates. The integral of the Gaussian-distributed noise
currents associated with the Josephson junctions is evaluated across
the multi-dimensional operating region of the circuit. Measurement
determines the value of a single parameter, effective noise bandwidth,
for each error mechanism, in a data-assisted approach. All this would
apply equally well to the other superconducting SFQ logic families.

The BER in combinational RQL gates such as the AND gate is governed by
the JTL-based error mechanisms, including spontaneous switching at
overbias and pulse annihilation at underbias. The effective noise
bandwidth is quite low for pulse annihilation errors, as it is based
on the clock rate, not the junction plasma frequency. The distributed
nature of the SFQ pulse means that multiple junctions are involved in
nucleating the error for spontaneous switching, so the effective
device size is that of multiple devices in parallel, in agreement with
\cite{klein1982thermal}. All this explains why the measured BER
extrapolates to $10^{-110}$, which is much lower than predicted by
simplistic estimates for isolated devices. In the long-junction limit
\cite{averin2006rapid}, JTL interconnect can be virtually dissipation
free while maintaining negligible BER, as ``the coupling between
mechanical and thermal modes vanishes''
\cite{fredkin1982conservative}. Ultimately this may lead reversible
logic gates as well \cite{herr2010method, osborn2020reversible}.

The clocked comparator of the RQL DFF gate is representative of the
other SFQ logic families, including RSFQ and the QFP. BER simulation
showed that the comparator produced a noise floor, at about $10^{-25}$
using junctions with 50\,$\mu$A critical current, as the comparator
does involve isolated junctions and higher noise bandwidths. This BER
is still quite low, suitable for petaflop-scale computing, but larger
junctions would be needed in order to maintain this level while taking
the parametric spread of individual junctions into account. For both
the DFF and the AND gate, extrapolated BER of $10^{-25}$ or better is
maintained across an AC power range of 3.6\,dB, equivalent to AC
amplitude margins of $\pm$20\%.

Overall we conclude that the BER in RQL circuits is sufficiently low
for large-scale applications, even with junction critical current
scaled down to 35\,$\mu$A minimum in the JTLs, as predicted
\cite{herr2011ultra}. BER performance of each gate in the library can
characterized and optimized by appropriate device sizing in the design
phase, using BER simulation as the primary tool.

\section*{Appendix}
We now describe some details of the numerical integration of the
Gaussian distribution across the operating region of the circuit. As
discussed above, mapping out the operating region is effectively to
find the global minimum by exhaustive search. If only the local
minimum were wanted, the downhill simplex method could be
used. Instead the entire operating region is mapped out with
simplexes. The algorithm uses an anneal to first search exhaustively,
and then to adaptively apply higher resolution to the regions of
interest.

The algorithm is as follows:
\begin{enumerate}
\item Calculate all $2N$ 1\,D margins and all $2^N$ corner margins
\item Make the initial simplexes, each using $N-1$ margins and $1$
  corner margin
\item Update the annealing weights
\item Find the greatest simplex based on the annealing weights
\item Bisect it through the longest segment by finding a new point on
  the operating boundary
\item Bisect all other simplexes for which this is the longest line
  segment
\item Calculate the solid angle subtended by each new simplex and
  update the Riemann sum
\item Until desired accuracy is achieved, goto (3)
\end{enumerate}

Three points require further explanation: A) simplex formation and
bisection, B) the algorithm is adaptive based on an annealing
schedule, and C) the solid angle subtended by each simplex must be
calculated.

\subsection{Bisection}
The first simplexes are formed in steps (1) and (2). The ``corner
margins'' consist of all search vectors for which each dimension has a
value of $\pm 1$. There is one corner vector centered in each orthant.
Many possibilities exist for subsequent iterations. Experimentation
indicates that keeping the aspect ratio of each simplex small is
important. This explains the conditional in step (6) of the algorithm.

The number of simplexes grows exponentially with $N$ and with
iteration number, and becomes a limiting factor above 8-D. Integration
in 7-D can create upwards of 500,000 simplexes; in 8-D, 5,000,000
simplexes. The problem is not storage per se, but the time spent
doing unordered searches in steps (4) and (5). Incremental sorting is
needed in step (4). Step (5) would benefit from an indexing system
whereby each point had a field indicating its simplexes. Alternately,
the simplexes could be organized by orthant. in this case, the signs
of the coordinates for each point could be used as the index. The
unordered searches in the current implementation limit us to about
8\,D.

\subsection{Anneal}
Each term $\Omega_s\langle\mbox{Q}_s\rangle$ in the Riemann sum is the
product of the solid angle subtended by the simplex and the average
value of Q for the points defining the simplex, The schedule computes
the annealing weight of each simplex as
$\Omega_s\langle\mbox{Q}_s\rangle^w$. The schedule moves the value of
$w$ from 0 to 1 incrementally. When $w=0$, the algorithm bisects the
simplex with greatest solid angle; when $w=1$, the simplex that is
most significant. In this way the algorithm starts by covering the
space uniformly, and gradually becomes more adaptive. The anneal could
be taken yet further by computing the variance among the values of the
points in each simplex, and weight the anneal accordingly. This
produced unstable results in our experimentation. However, we do
compute and sum these variances in order to estimate the error, as for
error estimates in Monte Carlo integration \cite{pressnumerical76}.

If the anneal proceeds too quickly, the danger is that it will find
some local minima, while missing others of more significance.  As
stated previously, $3^N$ points may be needed to cover the
space. However, we have found via a user-defined annealing parameter
that completing the anneal with $2.5^N$ points can achieve reliable
results.

\subsection{Solid Angles}
We need the solid angle subtended by each simplex, $\Omega_s$. (This
is the language of three dimensions. For higher dimensions, this
generalizes to the volume subtended on the unit hypersphere.) The
solid angle is normalized by the whole,
$\Omega_N=\pi^{N/2}/\Gamma(N/2+1)$, where $\Gamma$ is the gamma
function. The procedure for calculating the solid angle of a simplex
required some effort, as we invented our own.

First we calculate the volume, $V_s$, subtended by simplex $s$ on the
operating region boundary surface, by also including the point at the
origin. This volume derives from the matrix determinant. In $N$
dimensions
\[
  V_s=\left|\frac{1}{N!}\,\mbox{det}(v_1, v_2, ..., v_N)\right|
\]
where $(v_1, v_2, ..., v_N)$ are the unit vectors associated with the
vertices of simplex $s$. The points at the center of each facet of
simplex $s$ form a similar simplex. These points $(v'_1, v'_2, ..., v'_N)$
are calculated by excluding the points of the simplex one-by-one and
finding the center among the remaining points:
\begin{align*}
  v'_1 & = \langle v_2, v_3, ..., v_N\rangle \\
  v'_2 & = \langle v_1, v_3, ..., v_N\rangle \\
  v'_N & = \langle v_1, v_2, ..., v_{N-1} \rangle
\end{align*}
Here the angle brackets denote the mean value of the vector
coordinates. The vertices of the new simplex lie inside the unit
sphere. Renormalize them to unit vectors, and calculate the volume
of this new simplex, $V'_s$, again by including the point at the
origin. The method centers on the observation that the volume of the
new simplex is smaller by a factor of $V_f=1/(N-1)^{(N-1)}$, in the
limit of small angles. Scale inversely with $V_f$ to produce a new
solid angle estimate
\[
  V'_s=(N-1)^{(N-1)}\left|\frac{1}{N!}\,\mbox{det}(v'_1, v'_2, ..., v'_N)\right|
\]
This produces a much better estimate for the original solid
angle. Finally, we use both estimates to extrapolate to the best
estimate,
\[
  \Omega_s=(V'_s/V_s)^{F_N}V'_s
\]
where the first term produces a correction to $V'_s$. The fitting
parameter $F_N$ was found through experimentation.
\begin{align*}
  (F_3, F_4, ..., F_{10}) = (0&.0000, -0.0997, -0.1106, -0.1057, \\
                           -0&.0978, -0.0888, -0.0808, -0.0711)
\end{align*}
This method is effective because the correction to $V'_s$ is
small. Typical accuracy was better than 1\%, judged by summing the
solid angles $\Omega_s$ and comparing to the known value of
$\Omega_N$.

\begin{acknowledgments}

The authors acknowledge valuable discussions with Alexander Sirota,
Charles Wallace, and Henry Luo.

\end{acknowledgments}

\bibliography{ber_preprint}

\end{document}